\documentclass[preprint]{epl}

\usepackage{amsmath,amssymb}

\title{Concentration dependence of the transition temperature
in metallic spin glasses}
\shorttitle{$T_g$ vs. $c$ in metallic spin glasses}

\author{R. Serral Graci\`a\inst{1}\thanks{E-mail:
\email{rubeng@science.uva.nl}} \and Th. M. Nieuwenhuizen
\inst{1}\thanks{\phantom{E-mail: }\email{nieuwenh@science.uva.nl}}
\and I. V. Lerner\inst{2}
\thanks{\phantom{E-mail: }\email{ivl@th.ph.bham.ac.uk}}}
\shortauthor{R. Serral Graci\`a \etal}
\institute{
\inst{1}{University of Amsterdam, Institute for
Theoretical Physics -\\ Valckenierstraat 65, 1018XE
Amsterdam, The Netherlands} \\
\inst{2}{School of Physics and Astronomy, University of Birmingham -\\
Edgbaston, Birmingham B152TT, United Kingdom} }


\begin{document}

\maketitle

\begin{abstract}
The dependence of the transition temperature $T_g$ in terms 
of the concentration of magnetic impurities $c$ in spin glasses 
is explained on the basis of a screened \emph{RKKY} interaction. 
The two observed power laws, $T_g \sim c$ at low $c$ and 
$T_g \sim c^{2/3}$ for intermediate $c$, are described in a 
unified approach.
\end{abstract}

Metallic spin glasses such as \chem{Cu_{1-c}Mn_c},
\chem{Ag_{1-c}Mn_c}, \chem{Au_{1-c}Fe_c} are alloys formed
by magnetic impurities embedded in a noble metal. The
transition temperature of such materials depends on the
magnetic impurity concentration $c$. Different phenomena
dominate for different concentrations. The mutual
interaction between magnetic impurities is mediated by
electrons, the \emph{RKKY} interaction. It can be
understood as follows: the sea of electrons interact with
an impurity and the scattered wave interferes with the
incoming one. This creates a pattern of spin polarizations
that brings an oscillatory behaviour and a $1/r^3$ fall
off of the form $J(\bm{r})=A \cos(2 k_F \bm{r})/\bm{r}^3$
at $T=0$. For very low concentrations, less than $\approx
50\un{ppm}$, the interaction can be neglected and the
magnetic impurities act independently bringing the Kondo
effect. For larger concentrations though less than $\approx
10 \un{at.} \%$ the \emph{RKKY} interaction is the dominant
interaction and the spin glass phase appears. The
oscillatory nature of the interaction and the position
randomness of the impurities form a disordered magnet.
Above this concentration, the chance of having a
significant amount of impurities as first or second
neighbours is high, and consequently clusters are formed.
For even larger concentrations the percolation limit is
reached and ferromagnetism or antiferromagnetism,
depending on the type of magnetic impurities, appears. For
a review on these different regimes see Ref. \cite{mydosh}.

The spin glass region of concentrations (excluding the cluster region) 
exhibits two different behaviours. On one hand,
for concentrations lower than $\approx
1/2 \un{at.} \%$, the data points approach a linear curve,
$T_g \propto c$, while on the other hand,
for higher concentrations, up to $\approx10-15\un{at.} \%$ 
the fit turns to $T_g \propto c^{2/3}$. In this letter we 
explain these scaling laws in a unified treatment.

The key point lies in the fact that the \emph{RKKY}
interaction is not infinite ranged. At finite temperature,
phonons interact with the electron sea smearing out the
spin polarization pattern at large distances,
consequently the interaction is cut off at a length
$\Lambda_T$. It has the form

\begin{equation}
J(\bm{r})=A e^{-r/\Lambda_T} \cos(2 k_F \bm{r})/\bm{r}^3
\label{eq:rkky}
\end{equation}

For pure metals this cut off is the thermal coherence length
$\Lambda_T=\hbar v_F\beta /\pi\propto 1/T$. For disordered metals
where non-magnetic impurities exist alongside with magnetic ones,
the former start to play a role \cite{bergmann}: the electron wave
that scatters from the magnetic impurity diffuses around the
non-magnetic ones before it reaches another magnetic impurity. The
contributions of all such diffusive paths add coherently as long
as the distance between the magnetic impurities is smaller than
the thermal coherence length ${\Lambda}_T$. This leads to cutting
off the typical value of the interaction \cite{bergmann} as in eq.
(\ref{eq:rkky}), but with ${\Lambda}_T= (D \hbar \beta/\pi )^{1/2}
\propto 1/T^{1/2}$, where $D=v_F^2\tau/3$ is the diffusion
constant of an electron in a disordered metal and $\tau$ is the
mean free time for elastic scattering. Thus for $T\tau\lesssim1$
(assuming also that the elastic scattering dominates over the
inelastic, $\tau\lesssim\tau_{\text{inel}}$) the effective range
of the interaction for a disordered metal is shorter than for the
pure case. This defines a limit between the two situations that
can bring differences in the large concentration regime for
certain materials since the above inequalities may well be reached
in some cases.

Shegelski and Geldart \cite{shegelski} have derived
the range of an indirect-exchange
interaction in disordered metals which takes into 
account the \emph{RKKY} interaction and \emph{sd} 
scattering. They could well describe a wide range 
of experiments by fitting new length scales that 
appear in the problem. We shall not aim at fitting 
the data but to give the basic mechanism. We focus 
on the case where the concentration of magnetic 
impurities is changed with no other added impurities.
Our approach does not have adjustable parameters.

We use a Hamiltonian that takes into account the fact 
that the magnetic impurities, i.e. ``spins'', are 
present in some sites of the lattice and not in all 
of them \cite{nieuwenhuizen1, nieuwenhuizen2}

\begin{equation}
\mathcal{H}(s)=-\frac{1}{2} \sum_{\bm{r},\bm{r}'}
J(\bm{r}-\bm{r}') s_{\bm{r}} s_{\bm{r}'} c_{\bm{r}}
c_{\bm{r}'} - H \sum_{\bm{r}} s_{\bm{r}} c_{\bm{r}}
\end{equation}
where $s_{\bm{r}}$ represents the spin on site ${\bm{r}}$
and $c_{\bm{r}}=1,0$ whether a spin is present on site
${\bm{r}}$ or not. $J(\bm{r}-\bm{r}')$ is the \emph{RKKY}
interaction between sites $\bm{r}$ and $\bm{r}'$. This
model is used since it contains from the beginning a
dependence on the concentration via an average of the
$c_{\bm{r}}$. It is a random site problem, since the
randomness comes from the distribution of spins in a
lattice, and not a random bond problem where each bond
has a random strength, as is for example for the SK model.
This model can be solved in the low concentration limit
via a mean field approximation in the replica scheme
\cite{nieuwenhuizen1, nieuwenhuizen2}. The transition 
temperature was found to satisfy the following condition

\begin{equation}
c \sum_{\bm{r}} \tanh^2 [\beta J(\bm{r})]=1
\label{eq:sum}
\end{equation}
where $\bm{r}$ represents each of the sites of the lattice
since the factors $c_r$ have been taken in average.
Combining the \emph{RKKY} interaction in
eq. (\ref{eq:rkky}) with eq. (\ref{eq:sum}) we basically 
have an equation of the form

\begin{equation}
4 \pi c \int_0^{\Lambda_{T}} \upd r \; r^2 \tanh^2
\left[\frac{A \beta}{r^3}\right]=1
\label{eq:int}
\end{equation}
where the upper limit of the integral accounts for the
interaction cut off, and, as a first approximation, we can
consider that the oscillations of the cosine are not of
qualitative importance in the range where the hyperbolic tangent
squared has significant values. Eq. (\ref{eq:int}) gives the 
transition temperature $T_g$ for a given concentration $c$.
For small concentrations, in order to
satisfy the equation, the transition temperature has to be
small, which allows us to extend the upper limit of the
integral to infinity. It yields

\begin{equation}
4 \pi c \beta_g \int_0^{\infty} \upd x \; x^2 \tanh^2
\left[\frac{A}{x^3}\right] =1
\end{equation}
which leads to the expected result \cite{nieuwenhuizen1, nieuwenhuizen2}

\begin{equation}
T_g \propto c
\end{equation}

For larger concentrations, the transition temperature has
to decrease. Then the
upper limit remains finite and the $\tanh^2$ in eq. (\ref{eq:int})
can be approximated by $1$. We get

\begin{equation}
4 \pi c \int_0^{\Lambda_T} \upd r \; r^2 = c \;
\frac{4\pi}{3} \Lambda_T^3=1
\end{equation}

From the middle expression we see that the transition
takes place when there starts to be on the average more
than one impurity in the range of attraction of the
\emph{RKKY} interaction, as one might have expected. Since
for a disordered metal $\Lambda_T \propto T^{-1/2}$, the
scaling law then reads

\begin{equation}
T_g \propto c^{2/3}
\label{eq:nonlinear}
\end{equation}

We want to stress that the $2/3$ power law is intimately 
related to the fact that the system is considered to be 
a disordered metal, having $\Lambda_T \propto T^{-1/2}$. 
In this sense, we consider metals with a small amount 
of non-magnetic impurities (for a study on the effect of 
a variation on the concentration of non-magnetic
impurities see Refs. \cite{shegelski, vier}). For a pure 
metal, i.e.\ for $T_g\tau\gg1$, where the range of the 
\emph{RKKY} interaction is proportional to the inverse 
of the temperature, $\Lambda_T \propto T^{-1}$, the scaling 
law in eq. (\ref{eq:nonlinear}) becomes $T_g \propto 
c^{1/3}$. In the intermediate case, i.e. for $T_g\tau\sim1$, 
to consider the system as being disordered is not anymore 
a good approximation, a fact that may bring values of the
exponent $\phi$ lower than the $2/3$ predicted for the 
disordered case, i.e. for $T_g\tau\ll1$. An example of 
that is the case for \chem{Au_{1-c}Fe_c} where the exponent 
$\phi \approx0.58$ \cite{cannella}. 

We have explained here the two ``pure''
scaling laws $T_g \propto c$ and $T_g \propto c^{2/3}$ 
(or $T_g \propto c^{1/3}$ for pure metals)
corresponding to the canonical spin glass \cite{mydosh}.
These are the limiting situations and effective exponents 
found in experiments may lie between $1/3$ and $1$.
We suspect that in the experiments performed till now, a 
non-negligible amount of non-magnetic impurities were always 
present in the sample. We therefore propose to perform new refined 
experiments in order to test the presence of lower exponents $\phi$.

\acknowledgments
This work is part of the research programme of the
'Stichting voor Fundamenteel Onderzoek der Materie (FOM)',
which is financially supported by the 'Nederlandse
Organisatie voor Wetenschappelijk Onderzoek (NWO)'.
Th.M.N. acknowledges hospitality at the University of Birmingham.

\end{document}